\documentclass[runningheads,a4paper]{llncs}

\usepackage{amssymb}
\usepackage{graphicx}

\usepackage{amsmath,amssymb,amsfonts}
\usepackage{xcolor}
\usepackage{lscape}

\usepackage{url}
\newcommand{\keywords}[1]{\par\addvspace\baselineskip
\noindent\keywordname\enspace\ignorespaces#1}

\begin{document}

\mainmatter  

\title {Cross-dataset domain adaptation for the classification COVID-19 using chest computed tomography images}


%

%
%
\author{Ridha Ouni \inst{*}, \and
Haikel Alhichri 
}
\institute{ Department of Computer Engineering, College of Computer and Information Sciences, \\ 
King Saud University, Riyadh, Saudi Arabia, 11543. \\ 
\mailsa\\ 
}


\urldef{\mailsa}\path|{rouni@ksu.edu.sa,hhichri@ksu.edu.sa}|

%


%
%

\maketitle

\inst{*} Correspondng author: rouni@ksu.edu.sa

\begin{abstract}
Detecting COVID-19 patients using  Computed Tomography (CT) images of the lungs is an active area of research. Datasets of CT images from  COVID-19 patients are becoming available. Deep learning (DL) solutions and in particular Convolutional Neural Networks (CNN) have achieved impressive results for the classification of COVID-19 CT images, but only when the training and testing take place within the same dataset. Work on the cross-dataset problem is still limited and the achieved results are low. Our work tackles the cross-dataset problem through a Domain Adaptation (DA) technique with deep learning. Our proposed solution, COVID19-DANet, is based on pre-trained CNN backbone for feature extraction. For this task, we select the pre-trained Efficientnet-B3 CNN because it has achieved impressive classification accuracy in previous work. The backbone CNN is followed by a prototypical layer which is a concept borrowed from prototypical networks in few-shot learning (FSL). It computes a cosine distance between given samples and the class prototypes and then converts them to class probabilities using the Softmax function. To train the COVID19-DANet model, we propose a combined loss function that is composed of the standard cross-entropy loss for class discrimination and another entropy loss computed over the unlabelled target set only. This so-called unlabelled target entropy loss is minimized and maximized in an alternative fashion, to reach the two objectives of class discrimination and domain invariance. COVID19-DANet is tested under four cross-dataset scenarios using the SARS-CoV-2-CT and COVID19-CT datasets and has achieved encouraging results compared to recent work in the literature. 
\end{abstract}

\begin{sloppy}
\section{Introduction}
\label{intro}
On December 31, 2019, the World Health Organization (WHO) reported unknown cases of respiratory diseases that have spread in Wuhan, China~\cite{johnson_wuhan_2020}. The disease is identified as a new virus part of the Coronavirus family that can cause illnesses ranging from the common cold to more serious respiratory diseases. The new virus was later known as  COVID-19.  On January 30, 2020, due to the spread of this disease in China and many other parts of the world, it was classified as a public health emergency by the WHO~\cite{reportWHO2021}.  COVID-19 is a respiratory illness with symptoms similar to typical influenza. The transcription-polymerase chain reaction (RT-PCR) laboratory test is used as a reference tool for diagnosing COVID-19. In addition, X-rays and chest Computed Tomography (CT) scans are considered to be new information technology (IT) tools for COVID-19 diagnostics. What distinguishes the IT approach is its interpretability, which can help in fast decisions taken by doctors regarding COVID-19. 

CT is a painless and non-surgical imaging method characterized by speed and high accuracy. CT uses advanced X-ray technology to help detect many diseases and obtain detailed images of bones, internal tissues, and organs, where CT images give more details than traditional X-rays. Body parts absorb X-rays in unequal ways which allow the doctor to distinguish body parts and any changes due to disease~\cite{WhatisCT}.  Figure~\ref{CTimageExamplesFig} shows examples of CT images from COVID-19 patients.

\begin{figure}[!ht]
\begin{center}
     \includegraphics[width=6cm]{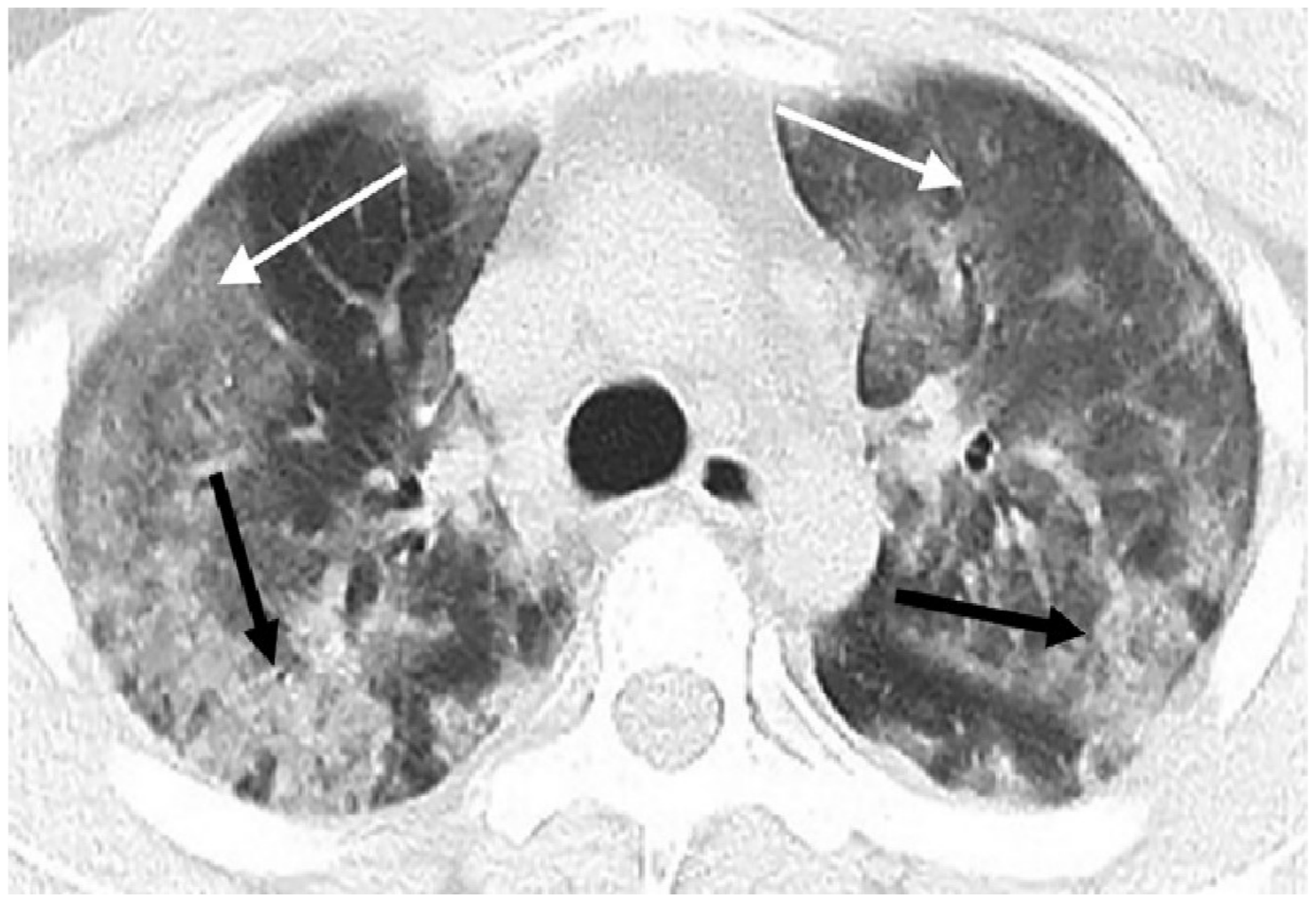}
     \includegraphics[width=6cm]{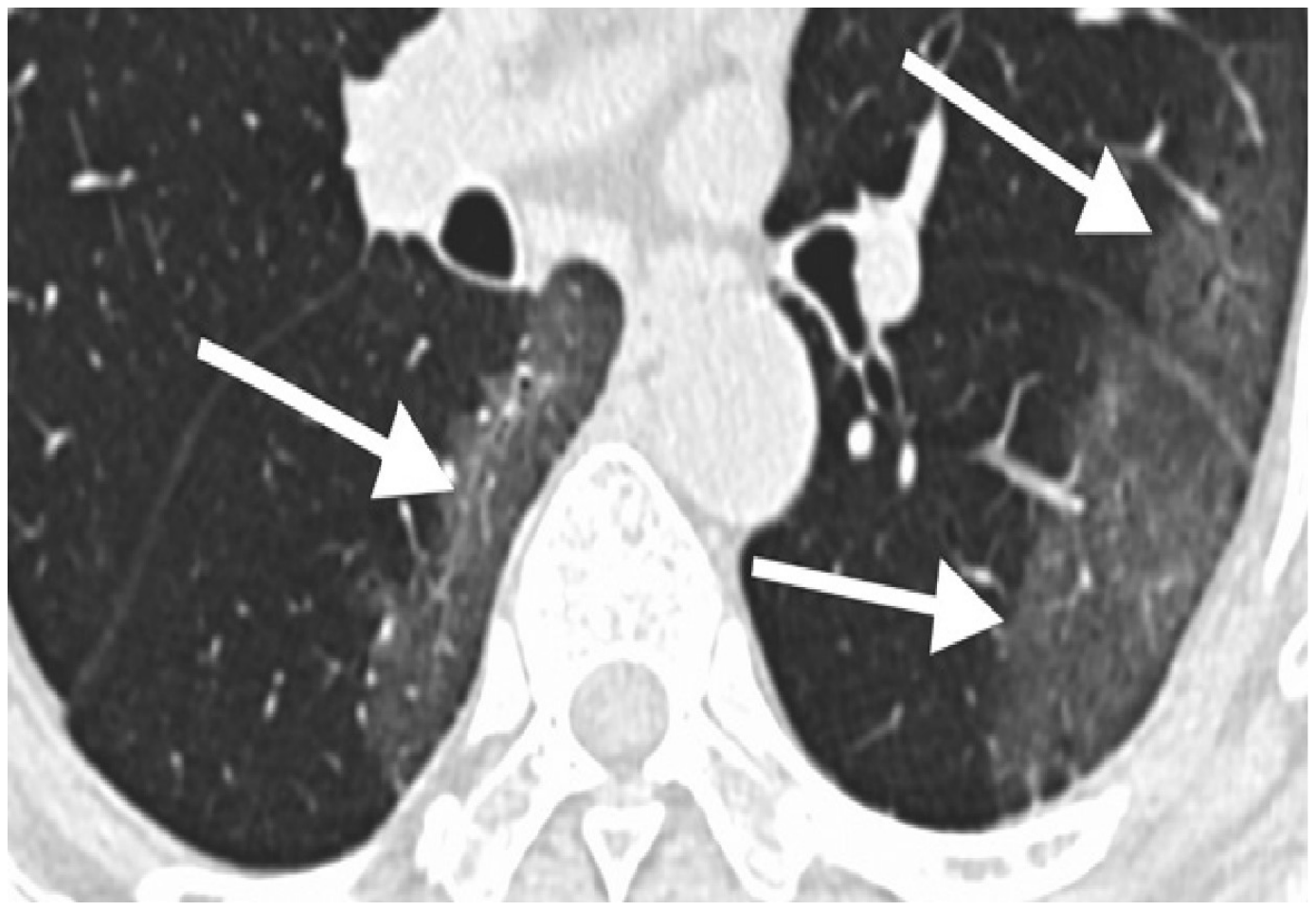}
\end{center}
\caption{CT scan shows ground-glass opacities in the lower lobes with a pronounced peripheral distribution (white arrows) and consolidative opacities (black arrows)~\cite{chung_ct_2020-1}   }
\label{CTimageExamplesFig}
\end{figure}

Ai et al.~\cite{ai_correlation_2020} report on 1014 patients who received both PCR and CT scans in Wuhan, China, during the epidemic. They found that 90\% of confirmed diagnostics of PCR had clear signs of COVID19 in chest CT scans appearing in the form of bilateral opacity. In another study, chest CT revealed bilateral opacity in the lung in 40 out of 41  patients (98\%) with COVID-19 in Wuhan~\cite{soares_sars-cov-2_2020}.  In addition, they show a high rate of appearance of ground glass opacity and uniformity with the round shape or occasional peripheral pulmonary distribution~\cite{soares_sars-cov-2_2020}. The probability of error in PCR tests, their limitations, and the length of time of their results, especially in areas affected by epidemic~\cite{xu_chest_2020}, as well as the low accuracy and sensitivity of X-rays in COVID-19 diagnosis, all made CT scans of the chest a very promising diagnostic tool for COVID19 \cite{bd_diagnosing_2020}.

Recently, advances in computer vision and Machine Learning (ML) have led to the emergence of a novel generation of techniques in computer-aided disease diagnosis (CAD)~\cite{BiswasKuppili2019}. In particular, Deep Learning (DL) in medical imaging has achieved outstanding performance in disease diagnosis and follow-up. DL  has proven its important role and efficiency in medical image processing including classification, detection and segmentation tasks~\cite{Alom_Taha_2019}.

In this Chapter, we propose a DL method for the diagnosis of COVID-19 disease using chest CT scans. In particular, we focus on the problem of removing the need for labeling large amounts of data to train DL models for  COVID-19 detection. In machine learning,  deep CNN models provide excellent results with large amounts of labeled data. However, this is neither reasonable nor practical. Ideally, our ultimate goal is to design a model that can provide good classification results for new datasets without the need for a large labeling effort. One solution is cross-dataset learning, where we transfer knowledge from one labeled dataset to another unlabeled dataset. This is also known as domain adaptation (DA) in which we develop learning models that can intelligently adapt from one source domain (dataset) to another target domain. To our knowledge, this is the first work that addresses the problem of DA in COVID-19 detection.

First, recall that the basic assumption of many machine learning algorithms is that training (source) and test (target) data come from the same distribution. However, in DA, the training and testing data come from different datasets taken under different circumstances which disproves the validity of this assumption. The distribution of data between source (training) and target (test) domains may change, causing low classification accuracy on the target data. The difference in distribution between different domains is still a very relevant problem among medical image datasets due to different image acquisition machines 
 and circumstances. As a result, there is increased interest in DA within the field of medical images to solve this problem and improve classification performance. But in the area of COVID-19 detection, the DA research is still very limited \cite{sinno_jialin_pan_survey_2010,guan_domain_2021}. In this objective, our proposed method is based on a new family of powerful CNN models called EfficientNet and on DA techniques to transfer knowledge from one domain to another. Our main contributions can be summarized in the following points:

\begin{itemize}
\item{  We present a DA method, called COVID19-DANet that can adapt a DL model from a source COVID-19 dataset to a target dataset. It uses the unlabeled samples from the target dataset to reduce the  data distribution shift between the source and target datasets. }
\item{  The proposed DA method uses EfficientNet-B3 CNN as a feature extractor and a classification layer inspired by prototypical networks from the few-shot learning area. }
\item{  Inspired by the semi-supervised learning methods proposed in the machine learning literature, COVID19-DANet uses the entropy of the output probabilities over the unlabelled target set as a loss function for reducing the distribution shift between domains.}
\end{itemize}

The remainder of the chapter is organized as follows. In Section~\ref{RelatedEWork}, we review some related work using CT images for  COVID-19 detection. We also survey some approaches based on DA learning, especially its field of application and typical setup. In Section~\ref{proposedMethod}, we present the  proposed DL-based DA model for COVID19 CT classification. Next, we present our experimental results in section~\ref{experimental} in terms of the most relevant assessment metrics. Finally, we outline our concluding remarks and future work suggestions in Section~\ref{conclusion}.

\section{Related work}
\label{RelatedEWork}

\subsection{Detection of COVID-19 using CT images}

The Center for Disease Control (CDC) in the United States of America has determined  the specific test for COVID-19 diagnostics is viral, while chest tomography or X-rays illustrates the features of COVID-19~\cite{bd_diagnosing_2020}. At present, RT-PCR is still the primary method of detecting COVID-19. But, each technology has its limitations and RT-PCR tests may not be available everywhere and all the time. With increasing incidents of COVID-19, especially in areas with a high epidemic severity, there may be a lack of availability of RT-PCR or high delay in the appearance of RT-PCR results~\cite{xu_chest_2020}. This  may negatively affect the spread of the disease~\cite{bd_diagnosing_2020}, in addition to the erroneous results of the initial RT-PCR that were not rare cases. The British Society for Chest Imaging confirmed the role of radiography in the diagnosis of COVID-19, especially in case of doubts about the diagnosis~\cite{xu_chest_2020}. CT of the chest was considered as an important diagnostic tool for the initial evaluation of COVID19.
Yan et al.~\cite{li_coronavirus_2020} conducted a study aimed at determining the error rate in which the radiologist may fail when doing a chest CT and identifying the features of COVID-19's CT and comparing them with the features of the CT of other viruses. This study involved 53 patients, including 51 patients who were diagnosed with COVID-19 infection and two patients with adenovirus infection. They noted an overlap in the results of CT of coronavirus and adenovirus.

CT is a standard tool in accelerating diagnosis for COVID-19 and proving the role of a skilled radiologist in the accuracy of diagnosis. A study carried out by Chunqin et al.~\cite{long_diagnosis_2020} suggested isolating patients who had abnormal chest CT results although the initial rRT-PCR results were negative and then repeat rRT-PCR to avoid misdiagnosis. In this study, 36 patients suspected of having COVID-19 underwent CT scan examination as well as a preliminary rRT-PCR examination. CT scans showed abnormal results for 35 patients and a normal image for one patient. However, when examined by rRT-PCR, the result was positive for 30 patients in the initial test, additional 3 patients in the second test, and 3 others in the third test. According to the previous study, the sensitivity of rRT-PCR was 83.3\% in the initial test, while the sensitivity in CT scans was 97.2\%.

Another study conducted an extensive review and analysis of chest CT and its accuracy in detecting coronavirus. Sixteen studies, covering 3186 patients, were divided into two groups according to the location of the study. They found that the most affected cities with coronavirus (Wuhan) had high sensitivity values for CT (97\%, 98\%, 99\%). While those values varied from 61\% to 98\% in cities other than Wuhan. After combining the shown information, the studies summarized abnormal chest CT features and patients who had positive chest CT. These studies showed initial false-negative RT-PCR in 36 patients, but initial chest CT showed positive in 31 out of 36 patients. The main results of this review are the effectiveness of CT in high epidemic areas, as well as the availability, convenience, speed, and its role in early diagnosis and control of the disease ~\cite{xu_chest_2020}.

The increasing rise in the number of patients infected with COVID-19 infection and the inadequacy of professional medical personnel have led to lower accuracy of the tests and thus lower the accuracy of the diagnosis of patients. Many studies, specifically DL-based approaches, have proven their active and accurate role in the diagnosis and classification of COVID-19 patients through CT scans of the lung. Yang et al. built an open-source COVID-CT dataset and then developed methods based on multi-task learning and self-supervised learning~\cite{yang_covid-ct-dataset_2020}. Soares et al. built and made available a large dataset SARS-COV-2 for CT, collected from hospitals in Brazil from real patients, and then introduced a new approach named explainable DL (xDNN) to identify COVID-19 using CT~\cite{angelov_explainable-by-design_2020}. In a similar work, Angelov et al.~\cite{soares_sars-cov-2_2020}, proposed an approach to detect COVID-19 via CT images depending on DL. Their approach provided high performance, as well as its ability to explain how the decision is made and its continued ability to learn from new data~\cite{soares_sars-cov-2_2020}. Xu et al. introduced an early screening model that automatically  detects COVID-19 from CT scans using DL techniques in order to distinguish COVID-19 from influenza-A viral pneumonia (IAVP) and healthy cases. They obtain promising diagnostic results~\cite{xu_deep_2020}. Li et al. proposed a framework based on DL, which they call COVNet, to distinguish COVID-19 from Community-Acquired Pneumonia (CAP) using CT scans~\cite{li_artificial_2020}. Wang et al. developed a DL algorithm to extract radiological graphical features from CT images to detect COVID-19~\cite{wang_deep_2020}. Jaiswal et al. introduced the COVID-19 classification model based on Deep Transfer Learning (DTL) using the pre-trained DenseNet201 CNN model. It has achieved a 97\% classification accuracy~\cite{jaiswal_classification_2020}.

Zheng et al. developed an algorithm based on DL weakly-supervised model that was trained on 3D CT images with patient-level labels (showing whether the patient has COVID-19 positive or negative). This study constitutes the first work based on weakly-supervised process that automatically detects COVID-19 over a large number of CT volumes. This model obtained a high accuracy of 90.1\% for identifying COVID-19 patients~\cite{zheng_deep_2020}. In another study,~\cite{hu_weakly_2020} presented a design proposal for weakly supervised DL using CT images for the automatic detection and classification of COVID-19 infection. The design was based on chest CT images obtained from multiple centers and multiple scanners. The proposed model allows for distinguishing COVID-19 cases from Community-Acquired Pneumonia (CAP) and Non-Pneumonia (NP). It demonstrates high classification accuracy as well as its ability to detect the exact location of the lesions~\cite{hu_weakly_2020}.
Harmon et al. introduced a series of DL algorithms, trained in a variety of nationalities and diversified in terms of CT scans of the chest to detect COVID-19. This study used COVID-19 CT scans from four hospitals across China, Italy, and Japan where they included a great variety of clinical timing and practice to obtain CT. These algorithms were evaluated on an independent test group (different from the training group) and achieved 90.8\% accuracy for classification of COVID-19 with sufficient generalization to other patient clusters/centers~\cite{harmon_artificial_2020}.
Song et al. proposed a DRE-Net architecture based on DL that aims to quickly and accurately diagnose COVID-19 using CT. The model proved to be able to distinguish between bacterial pneumonia and viral pneumonia (COVID-19) as well as to automatically extract the features of COVID-19. The proposed DL model achieved high performance and outperformed pre-trained models such as ResNet, DenseNet, and VGG16 in both the detection and classification of pneumonia~\cite{ying_deep_2020}.
Chen et al. developed a model based on DL to detect COVID-19 from other diseases using CT scans. For validation, the model achieved similar performance compared to the radiologist’s decisions~\cite{chen_deep_2020}. Gozes et al. introduced a DL system based on 2D and 3D models which are tested on international patients from the United States and China. The system detects the features of  COVID-19 from a suspected CT scan, as well as monitors the progression of the disease for each patient and assigns a "Corona score"~\cite{gozes_rapid_2020}.

Jin et al. built a CT scan dataset collected from three publicly available databases and three centers in China. They introduced an AI system based on a deep CNN model to detect COVID-19. The system showed good performance compared to the performance of radiologists with an accuracy of 94.98\%~\cite{jin_development_2020}.
Shi et al. put in place a method called an Infection Size Aware Random Forest (ISARF), where subjects were automatically divided into groups whose ranges vary according to the infected lesion sizes found in CT scans, in order to screen COVID-19 patients out of community-acquired pneumonia (CAP)~\cite{shi_large-scale_2021}. Bo Wang et al. designed and deployed an AI system for rapid and automatic detection of COVID-19 using CT. The system includes classification and segmentation of infection areas, which enhances the accuracy of diagnosis for doctors~\cite{wang_ai-assisted_2021}. Shuo Wang et al. provided a DL system for early prevention of COVID-19, focusing on abnormal areas of the lung. The system excels in distributing patients into groups based on the degree of risk, helping to identify high-risk patients in the absence of human support~\cite{wang_fully_2020}.
Fu et al. presented a framework based on AI for  the classification of COVID-19 and other common lung diseases using CT images~\cite{fu_deep_2020}. Maghdid et al. created a dataset consisting of CT and X-ray images and applied a DL algorithm using the AlexNet model to detect COVID-19~\cite{maghdid_diagnosing_2020}.  Gozes et al. have developed a system based on unsupervised DL algorithms to detect COVID-19 from CT scans. The system clusters, segments, and classifies images in order to determine the severity and progression of the disease~\cite{gozes_coronavirus_2020}. 

Alom et al. presented an improved approach using multi-task DL model to detect COVID-19 patients. Their approach was based on the Inception Residual Recurrent Convolutional Neural Network (IRRCNN) model for classification tasks as well as the NABLA-N network ($\nabla$ N-Net) for segmentation tasks, and then tested on CT and X-ray images~\cite{alom_covid_mtnet_2020}. Mobiny et al. proposed a new structure for a detail-oriented capsule network (DECAPS) to identify features in CT images for COVID-19 patients~\cite{mobiny_radiologist-level_2020}. Polsinelli et al have designed a light CNN to distinguish between CT images of COVID-19, community-acquired pneumonia (CAP) and/or other healthy cases. The proposed design relied on the SqueezeNet model with an accuracy of 83\%~\cite{polsinelli_light_2020}.
Sun et al. presented a proposed method based on the Deep Forest model for an Adaptive Feature Selection (AFS-DF) to differentiate between COVID-19 and CAP using CT~\cite{sun_adaptive_2020}.
Kang et al. diagnosed COVID-19 from CAP by proposing pipeline a latent multi-view representation learning in order to find integration between the different features of CT images that enhance diagnostic performance~\cite{kang_diagnosis_2020}.
Chen et al. proposed a DL algorithm to classify CT images for COVID-19 patients based on Few-Shot learning that used  few training samples~\cite{chen_momentum_2021}.
Lokwani et al. developed an extended version of the U-Net model characterized by a 2D segmentation task to identify areas of COVID-19 infection in CT scan~\cite{lokwani_automated_2020}.

Saeedi et al. introduced a CNN pipeline consisting of ResNet, Inception, MobileNet, and DenseNet in order to determine the most efficient diagnosis model~\cite{saeedi_novel_2020}. Then, they used SVM to classify CT scans into two classes (COVID-19, non-COVID-19). The proposed design was published to the public on the  Internet service for automatic detection of COVID-19 through CT images~\cite{saeedi_novel_2020}.
Hasan et al. designed a proposed DL network called Coronavirus Recognition Network (CVR-Net) that is tested on different datasets of CT and X-rays images~\cite{hasan_cvr-net_2020}.

Ozkaya et al. presented the only work that is not based on DL as they proposed a method to detect COVID-19 that extracts and  fuses handcrafted features and then uses the Support Vector Machine (SVM) to classify the processed data~\cite{ozkaya_coronavirus_2020}. 

Finally, in a recent interesting work, Silva et al.~\cite{silva_covid-19_2020} presented a vote-based technique for COVID-19 screening. In a patient-based split, they tested the proposed technique on two datasets: COVID-CT and SARS-COV-2-CT datasets. They are the only ones that presented a limited study on cross-dataset classification of CT images. The results found that in the best assessment scenario, the accuracy drops from 87.68\% to 56.16\% for the source and the target datasets respectively. The results showed that we are still far from a general solution for COVID-19 screening using CT scans.

\subsection{Domain adaptation learning}

DA is an approach employed to remove or reduce the need for labeling a lot of data for training. In DA methods, we assume that we have a source dataset that is labeled fully, and a target dataset that has a few labeled samples per class or none at all (i.e. fully unlabelled).
Depending on whether or not labelled data from the target dataset is used, DA can be unsupervised~\cite{ghifary_deep_2016,benjdira_unsupervised_2019} or semi-supervised~\cite{hoffman_simultaneous_2017,kushibar_supervised_2019,adayel_deep_2020,choudhary_advancing_2020,lasloum_ssdan_2021}. Specifically, if no labelled data is used from the target data, then we have an unsupervised domain adaptation (UDA) scenario. In this case, all we have is the unlabelled target data which can still be used to learn valuable information. In the semi-supervised domain adaptation (SSDA) scenario, only a few of the target data is class labeled. In addition, the remaining data from the target is used as an unlabelled set. In this chapter, we focus on the SSDA scenario because it provides better performance than UDA. Moreover, it is a practical scenario where minimal input is needed from an expert who will label a few (less than ten) images per class.

Most DA approaches try to find a feature space such that source and target data cannot be distinguished from each other~\cite{wilson_survey_2020}. The typical setup is illustrated by Figure~\ref{DAtypicalSetup}. This setup is inspired by Generative Adversarial Networks (GAN)~\cite{goodfellow_generative_2020} which is successfully adapted to solve the UDA problem~\cite{liu_coupled_2016,tzeng_adversarial_2017,sankaranarayanan_generate_2018,benjdira_unsupervised_2019}.

\begin{figure}[!ht]
\begin{center}
\includegraphics[width=12cm]{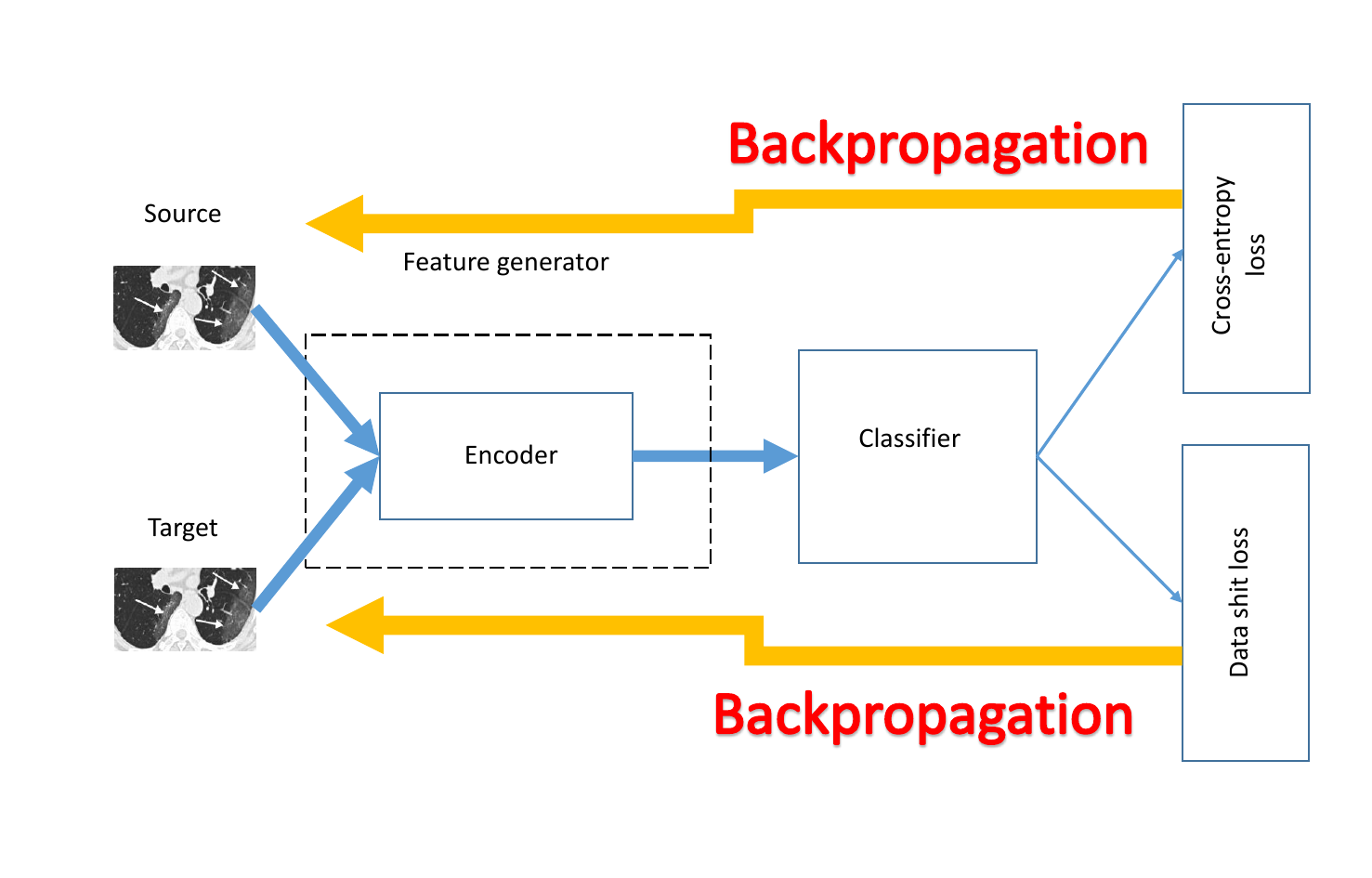} 
\end{center}
\caption{Domain Adaptation typical setup.     }
\label{DAtypicalSetup}       
\end{figure}

First, the encoder is used to generate features from the source and target domains. Then, the classifier module considers the feature outputs of the generator and predicts the different image classes. 
During the training, the generator is rewarded when the features encoded from the source and target domains cannot be distinguished from each other. We can accomplish this by minimizing a special loss function that measures the similarity between the source and target features. Therefore, the encoder will eventually map the source and target images to a common feature space where they are indistinguishable. On the other hand, the encoder is  rewarded when the classifier module predicts the classes correctly. We accomplish this by minimizing the typical cross-entropy loss. During training, the two loss functions are minimized in an alternating fashion. In one step, the classifier is frozen and the encoder is rewarded if the model cannot distinguish between source and target features. In the following step, the encoder is frozen and the classifier is trained to classify images into different classes. Using this alternating training process, this model learns to remove the data shift between source and target domains (by mapping them to the same feature space), and at the same time, it learns discriminative features with respect to the image classes.

In this chapter, the EfficientNet CNN model is used as an encoder to generate features from the source dataset as well as the target dataset. We use a special classifier module that is adopted from the few-shot learning field (introduced in the next section), due to the limited number of labelled samples in the target dataset. We also employ a loss function adopted from SSL as an entropy function computed over the unlabeled data.

\subsection{Few Shot Learning}
\label{FS_learning}

Learning to classify instances of objects that belong to new categories, while training on just one or very few examples, is a long-standing challenge in modern computer vision. This problem is generally referred to as few shot learning (FSL)~\cite{chen_closer_2019}. Modern DL-based methods need huge amounts of labeled data samples to work well, whereas, a child can rapidly comprehend new visual concepts and recognize objects from a newly learned category given very few examples. Using only a few shots for training, modern DL-based methods tend to suffer from severe over-fitting and provide bad performance. Another challenge of FSL resides in that the model must also identify new classes that are never learned during the training phase. In other words, the model should be trained on $K$ samples from a set of $N$ classes, so that later on it can classify any samples from another unseen set of $N$ classes. However, unlike DA, in FSL the unseen classes are coming from the same domain or dataset. 

FSL is related to our work because we will use a few samples from the target dataset to guide the DA training process. Thus, we can benefit from some techniques used in FSL to deal with the limited number of labeled samples in the training set. For example, we consider using an L2 normalization on the features (output of the encoder) prior to the last linear layer and a temperature parameter $T$~\cite{chen_closer_2019}. To increase the confidence of the output, networks can try to increase the norm of features. But, increasing the norm does not change the direction of the vectors, so this does not necessarily increase the between-class variance. This problem can be solved using L2 normalized feature vectors. To increase the confidence of the output, the network focuses on bringing the direction of features from the same class closer together and isolating different classes. This technique has proven effective with FSL~\cite{chen_closer_2019}, which is the main reason we adopted it in our work.

\section{Proposed DL model for domain adaptation}
\label{proposedMethod}

Figure~\ref{fig3:system_overview} gives an overview of the DL solution proposed to solve the domain adaption problem. It is composed of two modules; an encoder and a classifier.

\begin{figure}[!ht]
\begin{center}
\includegraphics[width=12cm]{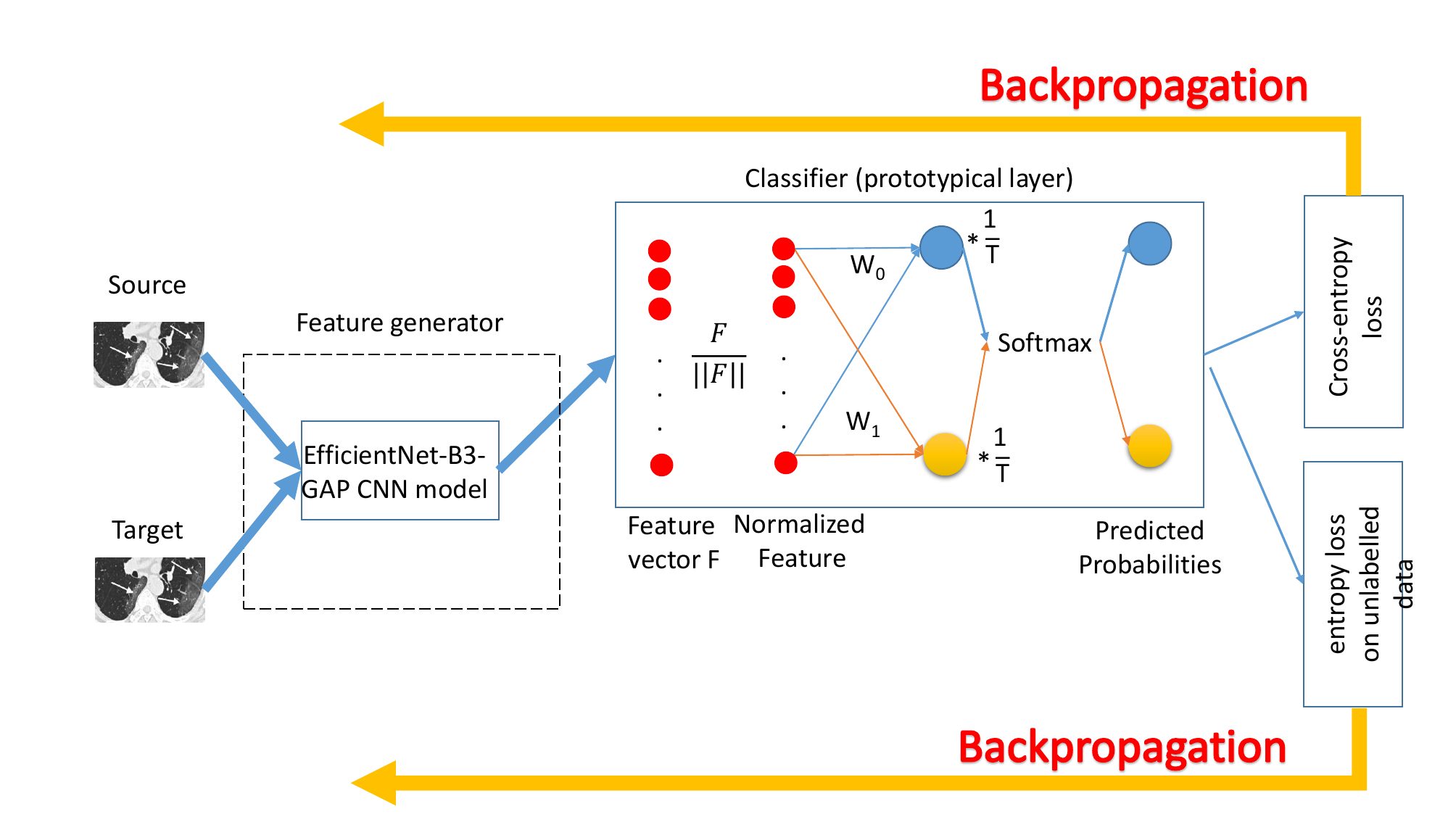} 
\end{center}
\caption{Proposed COVID19-DANet model for DA in COVID19 CT image classification.     }
\label{fig3:system_overview}       
\end{figure}

We make use of the EfficientNet-B3 model~\cite{tan_efficientnet_2019} as a feature encoder. EfficientNet-B3 belongs to a family of CNN models called EffecientNet models. Compared to the previous CNN models, EffecientNet models have achieved higher performance on the ImageNet dataset (which includes more than 14 million images), given a similar amount of network parameters.
The authors in~\cite{tan_efficientnet_2019} defined eight models in total ranging from the smallest EfficientNet-B0 to the largest EfficientNet-B7. We selected the EfficientNet-B3 model since it has shown significant classification performance despite having a reasonable small size~\cite{Alhichri_asmaa_2020,silva_covid-19_2020,alquzi_detection_2022}. The model size represents an important characteristic due to the limited computational resources available for this research work.

The classifier is based on prototypical networks used in the few-shot learning field \cite{NIPS2017_cb8da676,li_revisiting_2020}. The reason behind selecting this network is due to its effective performance with a small number of labeled samples. The classification module is modeled as a fully connected layer with a size equal to the number of classes (two in our case). The output of the classification layer is computed as follows:

\begin{equation}
\label{eq_similarity_layer}
p(x) = Softmax  \left(    \frac{1}{\tau}   W^T  .
\frac{   F_\theta(x)   }{   ||F_\theta (x)||     } 
\right)
\end{equation}

Where $F_\theta$ is the feature vectors and the operator $||.||$  denotes the L2 norm of the vector. $W$ represents the weights of the fully connected layer, and $.^T$  is just the transpose operator. $\tau$ is a scaling parameter called the temperature parameter in the jargon of few-shot learning. Finally, the results of this layer are passed through a Softmax activation function to convert the neuron outputs into probabilities (class predictions).

The division by the L2 norm is a normalization step. In practice, this step is very important due to its benefit in comparing feature vectors using a similarity metric, such as Euclidean distance. In the Equation~(\ref{eq_similarity_layer}), the term $(W^T . F_\theta (x))/(||F_\theta (x)|| )$ computes a similarity metric between $W$ and the normalized feature vectors $ F_\theta (x)/||F_\theta (x)|| $ of the sample x.  $W$ is actually composed of two vectors, $W_{0}$ and $W_{1}$, because we have two classes. Consequently, we can think of them as representing something like prototype feature vectors for each class that are learned during the training. Hence, the name "prototypical networks" is given to these types of networks in the field of few-shot learning. 
Thus, in this model, the class predictions are produced based on the similarity to the class prototypes. The term $(W^T . F_\theta (x))/(||F_\theta (x)|| )$ computes a cosine distance between the feature vectors of the sample $x$ and the two prototypes in $W$. Then, these distances are converted into probabilities using the Softmax activation function.

\subsection{Model optimization}
\label{model_optimization}

In our solution, we assume that the source dataset is fully labeled, whereas the target dataset includes only a few (less than 10) labeled samples per class. First, there is a high imbalance between the labeled samples coming from the source and training sets. This will make the model biased towards the source dataset. Thus, the model will be able to classify the source data well but not the target data. One simple solution is to make sure the training batches are balanced.

Recall that the limited amount of memory available in the Graphical Processing Unit (GPU), which we are using to speed up training, puts a maximum on the number of images we can use in parallel. Thus, when training deep neural networks, data is divided into batches of size $B$ because of these memory constraints. To insure balanced training, we make sure that an equal number $(B/2)$ of images comes from the source and target datasets in each batch.

Obviously, since the labeled source dataset is large, the sample images selected from it will vary much more than the sample images selected from the target dataset. For example, if the number of labeled samples from the target dataset is $K=3$, then there are only six labeled samples from the target dataset to select from (ignoring augmentation). Therefore, if the batch size is $B=12$, for instance, then we will always use all of the six target samples in every batch. On the other hand, the six samples coming from the source dataset will change randomly in every batch. 
We also randomly selected a few samples from the target dataset as a validation set. The validation set is not necessary, but it is useful as we can use it to monitor the model during the training process and decide when to stop. The decision to stop will use a criterion based on the validation loss or accuracy values.

This model is optimized using two loss functions. The first function, named standard cross-entropy loss, is computed over the labeled images from both source and target datasets. The cross-entropy loss, denoted by $L_{ce}$, is computed as follows:
\begin{equation}
\label{eq:eq_cross_entropy_loss}
L_{ce} = \frac{1}{N_s + N_t}  \sum_{i=1}^{N_s + N_t} \sum_{k=1}^{C}
1_{hot}\left(  y_{ik}=k  \right)
ln\left[
Softmax  \left(    
\frac{ W^T   F_\theta(x)   }{   ||F_\theta (x)||     } 
\right)
\right]
\end{equation}

where $y_{ik}$  is the true label, and $1_{hot}$ is an indicator function that returns one if the included statement is true, otherwise, it returns a zero. Furthermore, $N_s$  and $N_t$  are the number of labeled samples from source and target sets respectively, and $C$  is the number of classes. The cross-entropy loss ensures predicting the correct class for the labeled sample images. However, this function does not address the data shift problem between the source and target domains. In fact, the model will be heavily biased towards the source dataset since the majority of labeled samples come from there as opposed to the target dataset. As described earlier, one way to address this problem is to use an equal number of samples from both the source and target dataset in each batch of samples during the training. Additional help comes from including the unlabelled target samples in the training. This produces a domain invariant model as well as reduces the bias towards the source dataset. 

Conditional entropy minimization is one such approach used in SSL to learn from unlabelled samples \cite{long_conditional_2018}. Inspired by their work \cite{long_conditional_2018}, we include a second SSL-based loss function that involves the unlabelled images from the target dataset. This loss function, $E_u$, is called unlabeled entropy and is defined as the entropy over the predicted probabilities for the unlabelled images from the target dataset as shown in equation~\ref{eq_unlabaled_entropy_loss}:
\begin{equation}
\label{eq_unlabaled_entropy_loss}
E_u  =  -\frac{1}{N_u}  \sum_{i=1}^{N_u} \sum_{k=1}^{C}
p(y==k|x)  
ln\left[     p(y==k|x)         \right]
\end{equation}
where $N_u$ is the number of unlabeled samples and $p(y==k|x)$ is the prediction probability of class $k$ for unlabeled element $x$. 

The unlabeled entropy loss  $E_u$ can improve both the domain invariance of the model as well as the inter-class separation. Recall that in information theory, the highest entropy is obtained when all events have equal probabilities (uniform distribution) as illustrated in Figure~\ref{fig:understanding_entropy}. Inversely, if some probabilities are close to zero while the rest are close to one, the entropy will be minimum.  

\begin{figure}[!ht]
\begin{center}
\includegraphics[width=10cm]{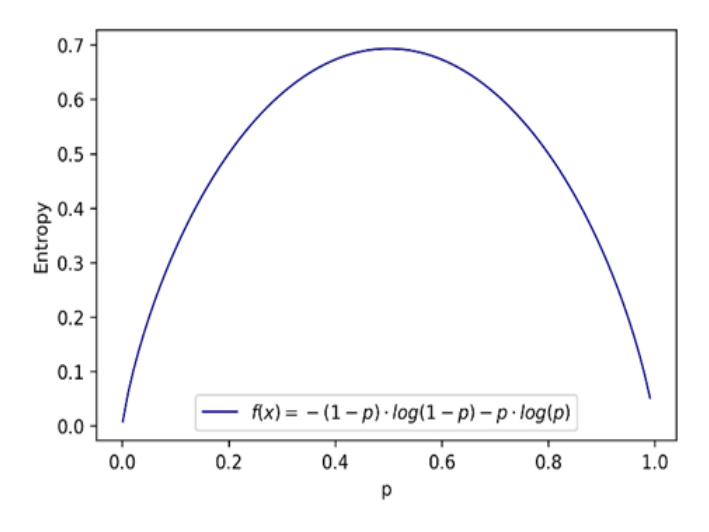}
\end{center}
\caption{Entropy as a function of probability. Entropy is highest when the two events have equal probability, otherwise, when one event has a probability of zero while the other has a probability of one, the entropy is equal to zero.}
\label{fig:understanding_entropy}
\end{figure}

We now illustrate the effect of the two loss functions on the feature vectors in Figure~\ref{fig:illustrating_loss_functions}. The cross-entropy loss guides the model to learn discriminative features that are separable into two classes. However, this only applies to the labeled samples, without covering the unlabeled samples. In fact, the unlabeled samples are wrongly classified as illustrated in this figure. To resolve this problem, we need to include the unlabeled samples and address the data shift problem. It turns out, we can reduce this shift by maximizing the unlabeled entropy loss $E_u$ with respect to the class prototypes \cite{li_learning_2021,zhao_domain_2020,lasloum_ssdan_2021}. 

\begin{figure}[!ht]
\begin{center}
\includegraphics[width=12cm]{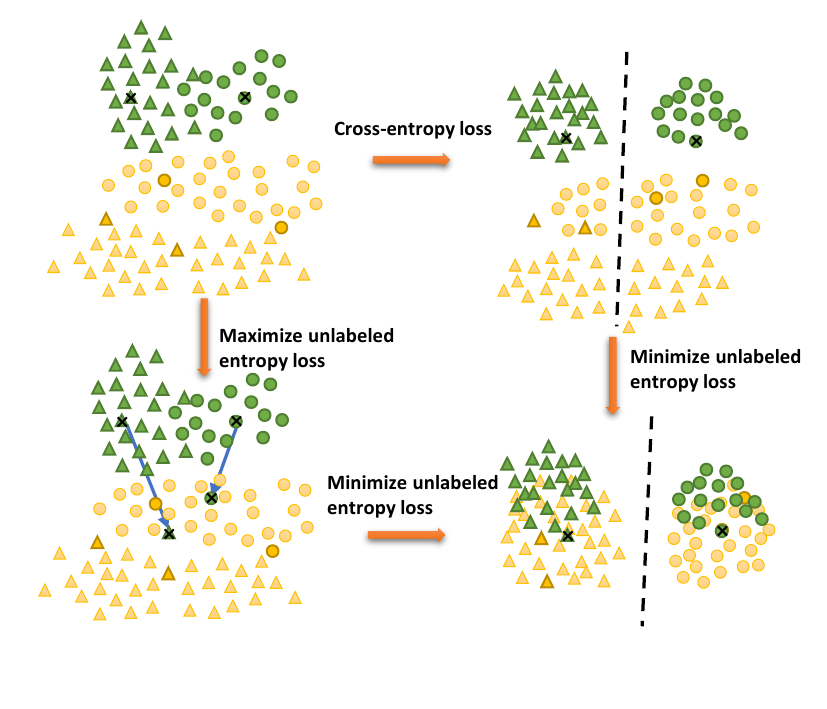}
\end{center}
\caption{Illustration of loss functions effect on learned feature vectors. The green color represents the source domain, while the target is in yellow. The light yellow color indicates the unlabeled samples. The $\times$ indicates the class prototype.}
\label{fig:illustrating_loss_functions}
\end{figure}

As illustrated in Figure~\ref{fig:illustrating_loss_functions}, maximizing the entropy $E_u$ results in uniform output probability distribution where the model learns a class prototype that is similar to all features of the unlabelled target samples.  Effectively, this idea brings the source and target features closer to each other reducing the data shift between the two domains.

On the other hand, when the unlabeled entropy loss $E_u$ is minimized, the probabilities for the unlabelled samples will be close to zero or one (as motivated by Figure~\ref{fig:understanding_entropy}), resulting in their features having smaller distances to the class prototypes. In other words, the features extracted for the unlabelled target samples become more compacted around the class prototypes $W_0$ and $W_1$, which means they are more discriminative.  

At this level, we have contradicting objectives for the loss $E_u$ which must be minimized and maximized at the same time. This issue is resolved by adversarial training inspired by the training approach in GAN theory~\cite{ganin_domain-adversarial_2017}. As result, two loss functions, defined in equation~\ref{eq_total_loss}, will be minimized alternatively.

\begin{equation}[!ht]
\begin{tabular}{cc}
$\psi_{H}$     & =  $L_{ce}  + \lambda E_{u}$ \\
$\psi_{C} $    & =  $L_{ce}   - \lambda E_{u}$
\end{tabular}
\label{eq_total_loss}
\end{equation}

How can these functions be minimized alternatively? When the first loss function is applied, the classifier $C$ is frozen, while when using the second function, the feature extractor $F$ is frozen. Note that minimizing $L_{ce}  - \lambda E_{u}$, results effectively on maximizing $E_u$ due to the negative sign. In this context, the $\lambda$ parameter is used to balance out the effect of the two losses. Several experiments are performed to study the effect of this parameter (section 4).

\section{Experimental results}
\label{experimental}

In this section, we present various CT image-based datasets. Then, we describe the pre-processing mechanism performed on the images and the experimental setup. Finally, we evaluate and analyze the experiment results.

\subsection{COVID-19 CT datasets}

\textbf{COVID19-CT}: This is the earliest open-source dataset prepared by Zhao et al.~\cite{zhao_covid-ct-dataset_2020}. It contains a total of 746 chest CT images, which are divided into two classes, namely COVID-19 and non-COVID-19. A pre-processed version of the dataset is available online~\cite{zhao_covid-ct-dataset_2020,yang_covid-ct-dataset_2020}. The dataset was created by collecting images from papers related to COVID-19 and was published in medRxiv, bioRxiv, NEJM, JAMA, Lancet, and other impact-full journals. These images were classified according to the figure captions describing the clinical findings in the papers. 349 CT images were labeled as COVID-19 and 397 CT images as non-COVID-19. The heights of these images range between 153 and 1853 pixels (average of 491 pixels), while their widths range between 124 and 1458 pixels (average of 383 pixels). 

\textbf{SARS-CoV-2-CT }: This dataset is considered as the largest dataset available for COVID19 CT scans collected from hospitals in São Paulo, Brazil. It consists of 2482 CT images, including 1252 images of 60 patients with COVID-19 and 1230 images of 60 patients with non-COVID-19 but with other pulmonary diseases~\cite{soares_sars-cov-2_2020,eduardo_sars-cov-2_2020}. In this dataset, samples are CT images printed with neither a standard size nor a standardization contrast.

\subsection{Data pre-processing}

We used the SARS-CoV-2-CT dataset~\cite{eduardo_sars-cov-2_2020} as an example to show the pre-processing steps in our work. Figure~\ref{fig:SampleCTscans_a} and Figure~\ref{fig:SampleCTscans_b} show sample CT images from the SARS-CoV-2-CT dataset. As indicated earlier, the images in this dataset have different sizes. Table~\ref{table_dataset_meta_information} shows the minimum and maximum width and height of the images in the dataset.

\begin{figure}[!ht]
\begin{center}
\includegraphics[width=12cm]{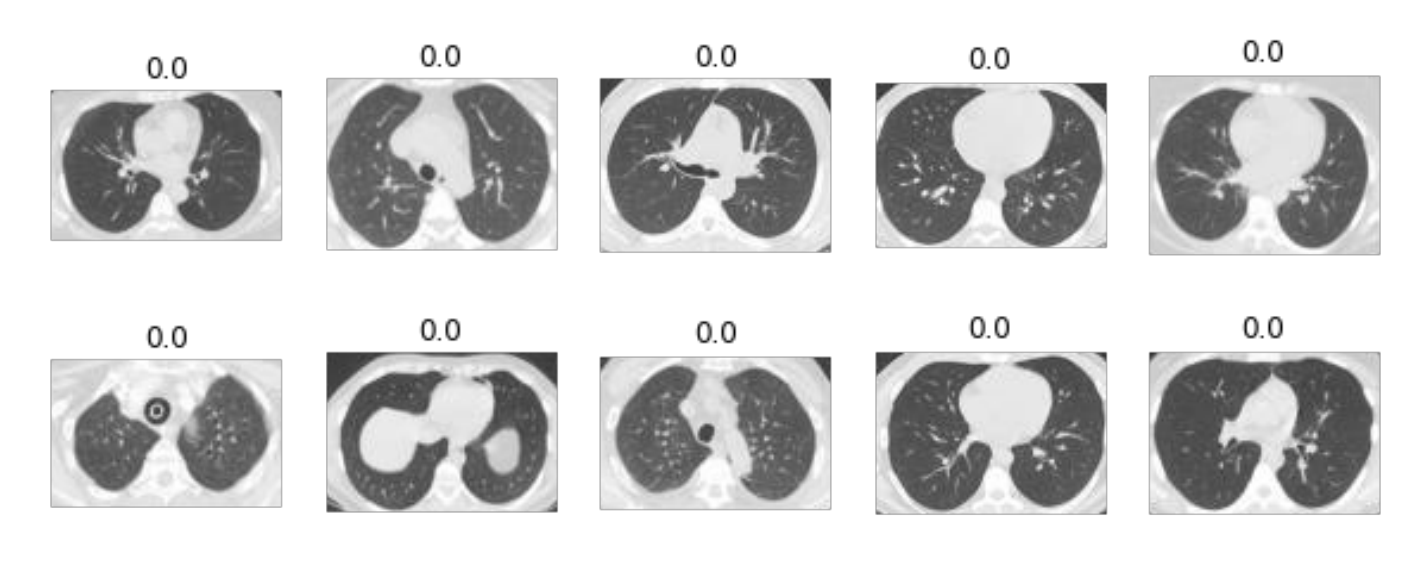} 
\end{center}
\caption{Sample of CT for different patients not infected with COVID-19. }
\label{fig:SampleCTscans_a}       
\end{figure}

\begin{figure}[!ht]
\begin{center}
\includegraphics[width=12cm]{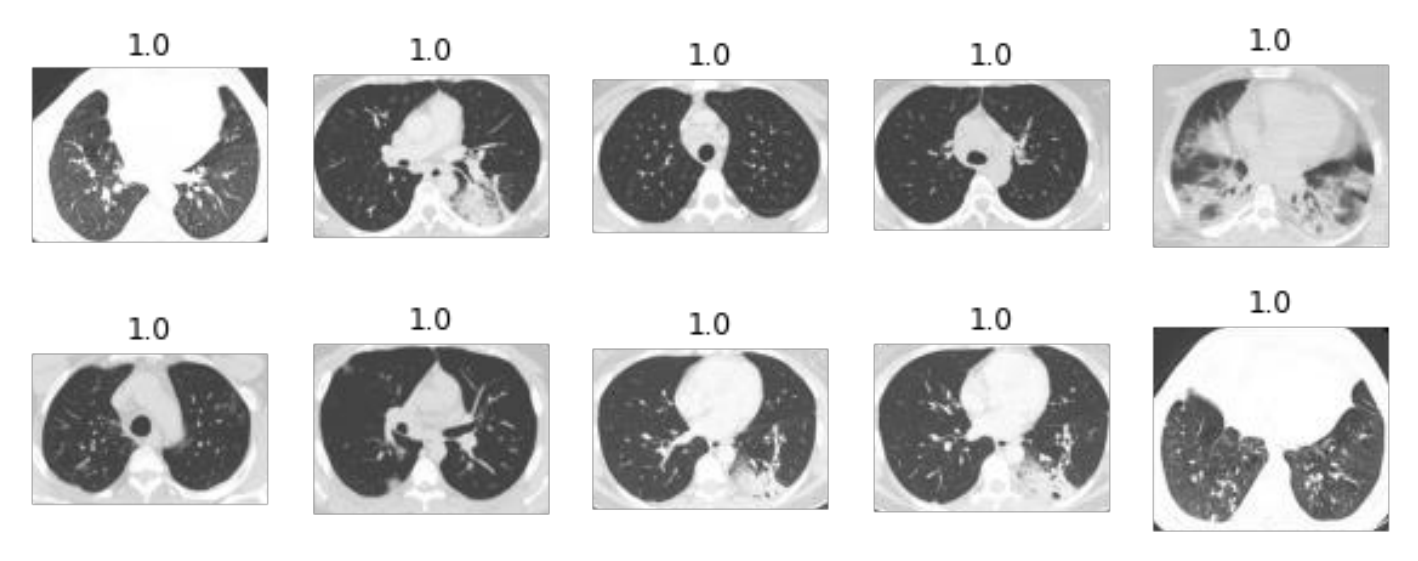} 
\end{center}
\caption{Sample of CT for different patients infected with COVID-19. }
\label{fig:SampleCTscans_b}       
\end{figure}

%
\begin{table}[!ht]
\caption{SARS-CoV-2-CT dataset meta information.}
\begin{center}
\label{table_dataset_meta_information}       
%
%
\begin{tabular} {p{3cm}p{2cm}p{2cm}p{2cm}p{2cm}}    
                 & \multicolumn{2}{c}{COVID-19} & \multicolumn{2}{c}{Non COVID-19} \\ \cline{2-5}
Number of images & \multicolumn{2}{c}{1252}     & \multicolumn{2}{c}{1229}         \\ \hline
Patients         & \multicolumn{2}{c}{60}       & \multicolumn{2}{c}{60}           \\ \hline
                 & Min           & Max          & Min             & Max            \\ \hline
Rows             & 123           & 408          & 119             & 416            \\ \hline
Columns          & 182           & 534          & 224             & 502           \\  \hline
\end{tabular}
\end{center}
\end{table}

To apply CNN models, the input images should have the same size. Early models only accept images of size $224\times 224$. However, new CNN models can accept other sizes as well. Thus, an important pre-processing step consists of resizing all images in the dataset to a common size. EfficientNets can exploit higher resolution input images due to their low computational cost in terms of latency and memory. In this work, the optimal image size can be defined as the middle ground between the minimum and maximum rows and columns (Table 1), so equal to $256\times 256$.

To deal with the variable image size problem, the first solution consists of resizing every image to the fixed target size. However, this option will distort the images significantly and affect their aspect ratio. As result, it will have a negative impact on the performance of the classification model. A second option proposes using a reference image size given by the maximum number of rows and columns, i.e. 534 (Table~\ref{table_dataset_meta_information}). Then, all images are downsized by a factor of $534/256 = 2.1$ and padding with zeros the images that end up with a size smaller than $256\times 256$. Albeit, this option performs better than the first one, still, the resolution of the smaller images will be significantly degraded. In addition, the padding with zero constitutes a large portion of these images with no benefit to the classification model. Due to these reasons, a third option is proposed where only the largest width of the image is used as a reference. 
Specifically, let us consider a sample image $i$ with a size of $W_i\times H_i$. This image should be resized by a factor of $\frac{256}{W_i}$. However, this resizing step scales the height to be $H_i\frac{256}{W_i}$, which may be greater or less than the target height (256). If the new height is smaller than 256, we simply pad zeros in equal amounts at the top and bottom, as shown in Figure~\ref{fig5:ImagePreprocessing}.

\begin{figure}[!ht]
\begin{center}
\includegraphics[width=12cm]{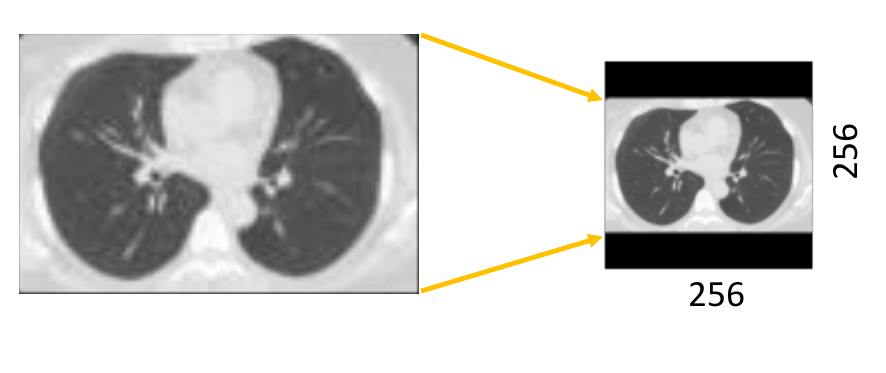} 
\end{center}
\caption{Example of preprocessing a sample CT scan image including the resizing and padding with zeros. }
\label{fig5:ImagePreprocessing}       
\end{figure}

Figure~\ref{fig6a} and Figure~\ref{fig6b} show the result of the pre-processing step on some images of the COVID19-CT dataset. To do this, if the new height is greater than 256, then we crop the image equally from the top and bottom to make its height 256. However, since most CT images are landscape, this case does not occur often.  In fact, in the whole SARS-CoV-2-CT dataset, this case only occurred ten times. Figure~\ref{fig7:badPreprocessingCases}(a) shows all these cases and as can be seen the amount to be cropped (outside the red box of size $256\times 256$) is always insignificant. The same pre-processing has been performed on the COVID19-CT dataset and resulted in similar observations. In this dataset, only two samples out of 812 experienced minor cropping, as shown in Figure~\ref{fig7:badPreprocessingCases}(b).

\begin{figure}[!ht]
\begin{center}
\includegraphics[width=12cm]{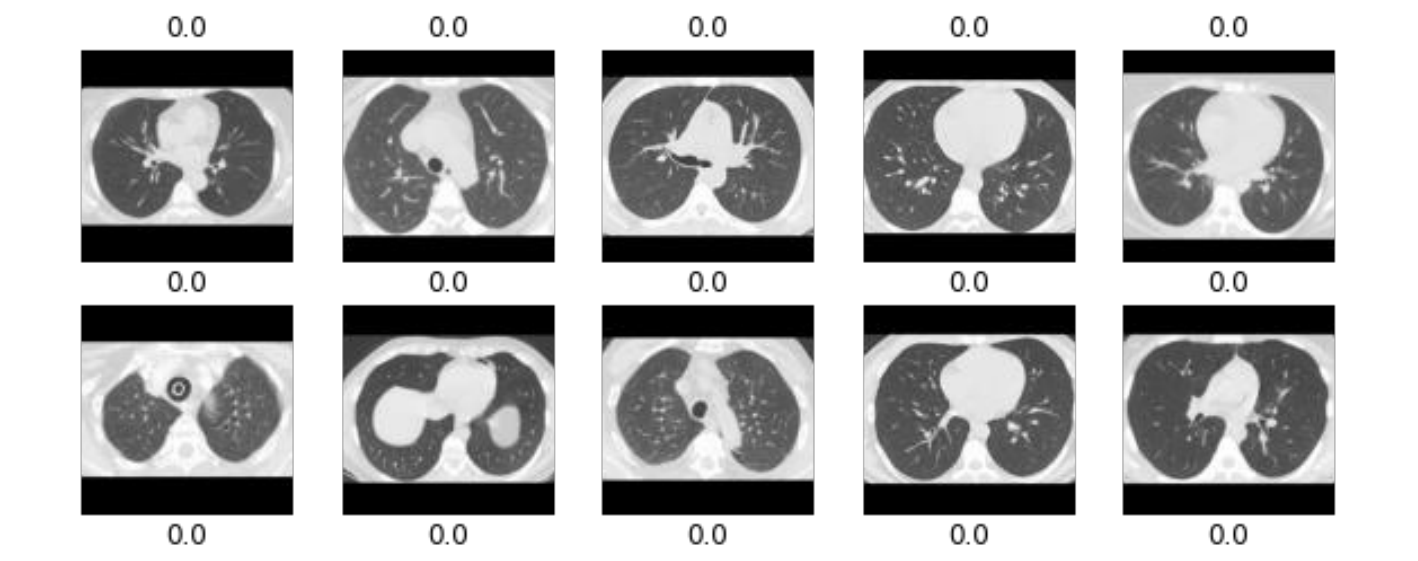} 
\end{center}
\caption{Sample of CT scan Non-COVID images after pre-processing. }
\label{fig6a}       
\end{figure}

\begin{figure}[!ht]
\begin{center}
\includegraphics[width=12cm]{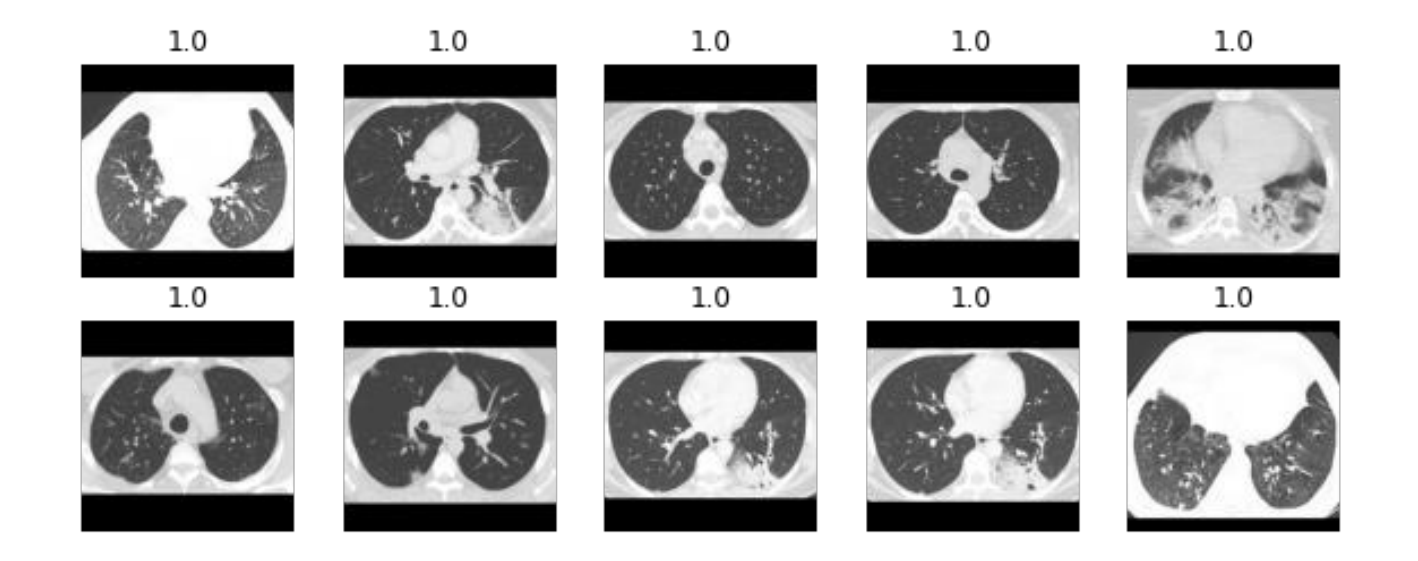} 
\end{center}
\caption{Sample of CT scan COVID images  after pre-processing. }
\label{fig6b}       
\end{figure}

\begin{figure}[!ht]
\begin{center}
\includegraphics[width=11cm]{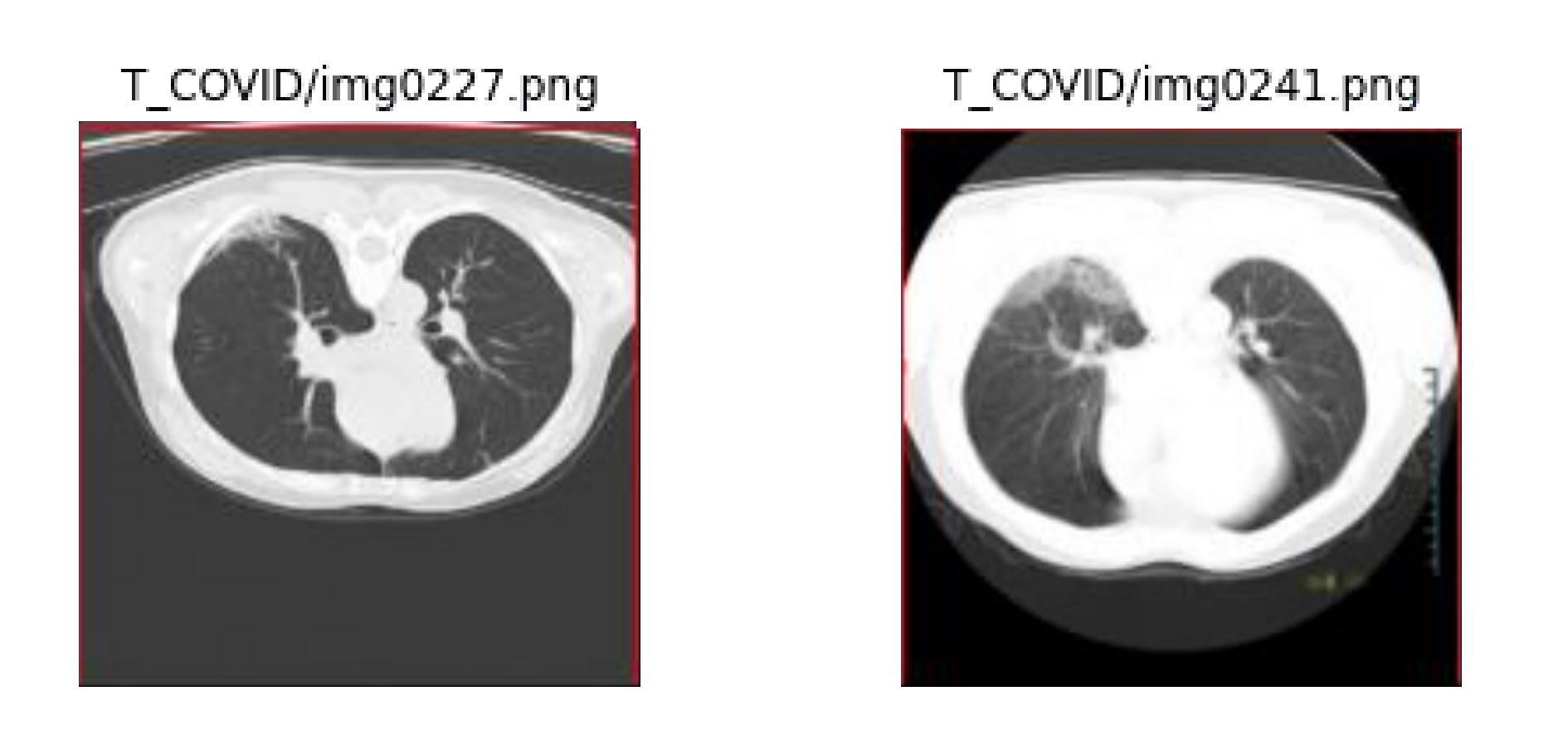} \\
\includegraphics[width=11cm]{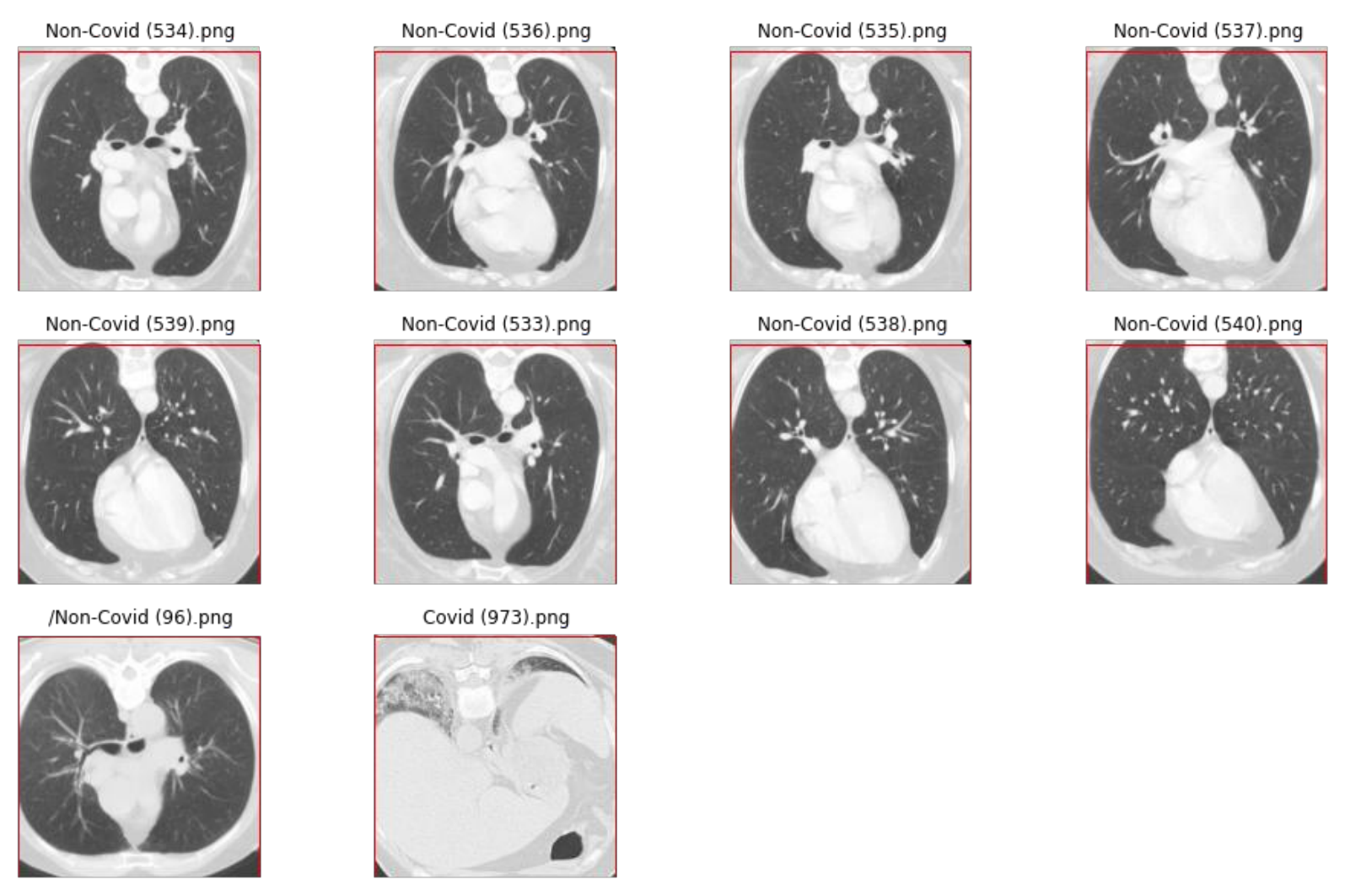} 
\end{center}
\caption{CT scan images that needed minor cropping during the proposed preprocessing. (a) images from the COVID19-CT dataset. (b) images from the SARS-CoV-2-CT dataset. }
\label{fig7:badPreprocessingCases}       
\end{figure}

\subsection{Assessment metrics}

The Accuracy assessment metric $A$ is used to evaluate the performance of the proposed approach compared to other related work. Accuracy measures the correctly recognized cases, and is calculated as:

\begin{equation}
\label{eq:Accuracy_metric}
A =  \frac{(TP+TN) }{(TP+TN+FP+FN)}
\end{equation}

$TP$ (or True Positive) represents the number of positive COVID-19 patients that are correctly identified. TN (or True Negative) represents the number of negative COVID-19 patients that are correctly identified. $FP$ (or False Positive) represents the number of negative COVID-19 patients who have other lung diseases, which are not recognized but have been identified as positive COVID-19. $FN$ (or False Negative) represents the number of positive COVID-19 patients that are identified as negative cases.

\subsection{Experimental setup}

Figure~\ref{fig8:SetupForDA} illustrates how the data are set up for this experiment. All labeled samples from the source dataset are used for training the DA model. In addition, $K$ labeled samples from each class of the target dataset are also used during this training.

\begin{figure}[!ht]
\begin{center}
\includegraphics[width=12cm]{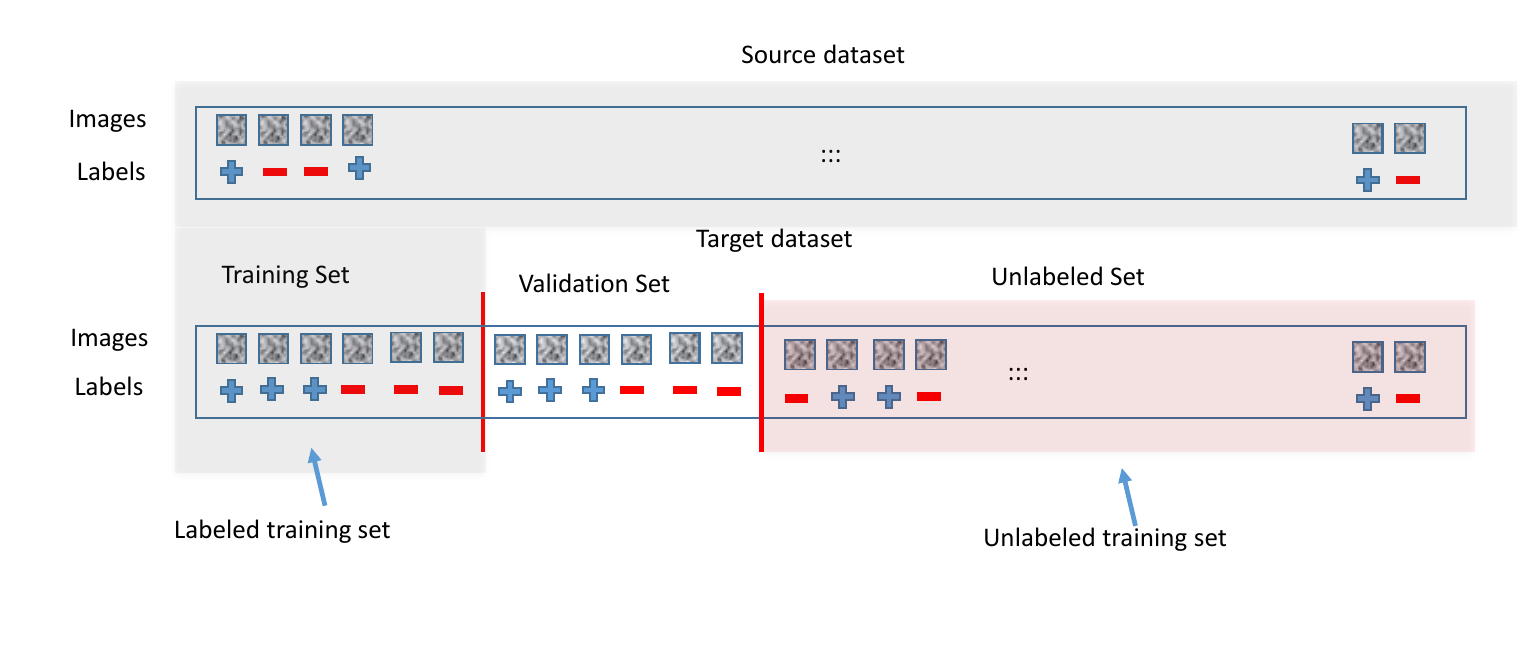} 
\end{center}
\caption{Setup for the cross-dataset domain adaptation experiments. }
\label{fig8:SetupForDA}       
\end{figure}

 The labeled samples from the source and target datasets are used to compute the cross-entropy loss as given in Equation~\ref{eq:eq_cross_entropy_loss}. Another $K$ labeled samples from each class of the target dataset are used as a validation set to monitor the training and save the model achieving the best validation accuracy. Finally, the remaining samples of the target dataset are used as the unlabelled set to compute the entropy loss function based on Equation~\ref{eq_unlabaled_entropy_loss}. 

Our proposed approach was developed using Python and Pytorch Machine Learning library. The network was trained using the Adam Optimizer with a learning rate $lr$ that decays from an initial value following Equation~\ref{eq:leraningRateScheduler} and depends on the batch number $b$~\cite{saito_semi-supervised_2019}:
\begin{equation}
\label{eq:leraningRateScheduler}
lr = lr \times (1 + \nu \times b)^{-p}
\end{equation}

Where the $\nu$ and $p$ parameters are set to the following default values of $0.001$ and $0.75$ respectively. The batch size for the experiment had to be computed carefully. During the training, each batch of labeled samples is composed of an equal number of samples from the source and target datasets. Large batch size is not always possible since (1) the number of labeled samples per class from the target dataset is limited ($3$, $5$, or $10$ in our experiments) and (2) only two classes (COVID versus Non-COVID) are available. In fact, if the number of labeled samples per class is three, then we only have six labeled target samples and the maximum possible batch size is $12$.  

\subsection{Data augmentation}

Data augmentation is a process that allows increasing the training set by applying transformations that do not affect the semantic information in the image significantly. For instance, CT scans can be flipped horizontally without affecting their size, content, and semantics. In this work, we used two transformations, namely horizontal flip and scaling. Obviously, after scaling, some information in the image might be lost. To reduce the loss, CT scans are scaled up or down by 20\%. These transformations do not introduce significant changes in the images and physicians can easily use them for diagnosis.
Data augmentation is applied to the labeled target dataset. This dataset is very small because we are only including $K$ shots from each class. Therefore, data augmentation constitutes an interesting process that increases the number of samples and improves the model performance. However, the source dataset has abundant samples, and augmenting it with more samples will make the model even more biased towards it. Thus, the source dataset did not incur any augmentation.

\subsection{Results and comparison to previous work }

Different DA scenarios are used to study the impact of our proposed algorithm for cross-dataset classification. Four DA scenarios, using COVID19-CT and SARS-CoV-2-CT datasets, are created based on the approach in \cite{silva_covid-19_2020}. The COVID19-CT dataset is already divided into training and testing sets, named COVID19-CT-train and COVID19-CT-test, while COVID19-CT-train-test denotes the full combined set. Table~\ref{table_5_DA_scenarios} defines four DA scenarios based on these sets.

%
\begin{table}[!ht]
\caption{Domain adaptation scenarios.}
\label{table_5_DA_scenarios}       
%
%
\begin{center}
\begin{tabular} {p{2cm}p{4cm}p{4cm}}    

\textbf{Scenario} & \textbf{Source  dataset} & \textbf{Target dataset} \\ \hline
Scenario 1          & SARS-CoV-2-CT             & COVID19-CT-test           \\ \hline
Scenario 2         & SARS-CoV-2-CT             & COVID19-CT-   train       \\ \hline
Scenario 3            & SARS-CoV-2-CT             & COVID19-CT-train-test     \\ \hline
Scenario 4             & COVID19-CT-train-test     & COVID19-CT-train \\  \hline
\end{tabular}
\end{center}
\end{table}

The first set of experiments employs the first scenario and investigates the performance under different parameter settings. The number of samples per class $K$ is set to three, so the total number of labeled target samples is only six and the batch size is fixed to 12. Let us recall here that half of each training batch must be from the target dataset. The learning rate is set to 0.0001 and decays according to equation (\ref{eq:leraningRateScheduler}). 

In these experiments, the epoch parameter is not used due to its dependency on the number of batches in the dataset. During the training, different sets are used namely the source, labeled target, and unlabelled target sets, which have drastically different numbers of batches. Therefore, it becomes difficult how to define one epoch of training. To resolve this problem, the number of batches is considered as the main training parameter. Here, we train the model for 400 batches and observed that the network loss is relatively stable at that point. 

During training, the network loss and accuracy are computed at the end of every batch. For the accuracy curves, we decided to only monitor the accuracy of both the validation set and test set (no significance for the training set accuracy), in order to observe the relationship between validation and test accuracy.  Computing the test accuracy of every batch is time-consuming. Subsequently, the computation is only performed at regular intervals. During this process, the model providing the highest validation accuracy is maintained in order to evaluate the classification accuracy over the whole target set. 

Several trials are performed to study the effect of the two most important parameters $T$ and $\lambda$ given in Equations (\ref{eq_similarity_layer}) and (\ref{eq_total_loss}), respectively. Both parameters are assigned the following values: 0.001, 0.005, 0.01, 0.1, 0.2, 0.5, and 1.0. As result, Figure~\ref{fig12}(a) shows the target accuracy obtained for $\lambda$ parameter variation, whereas Figure~\ref{fig12}(b) corresponds to $T$ parameter variation. Figure~\ref{fig12}) shows that the best accuracy is achieved for $\lambda =0.1$  and $T=0.05$. This process determines the optimal values of the input parameters for the remaining experiments.

\begin{figure}[!ht]
\begin{center}
\includegraphics[width=10cm]{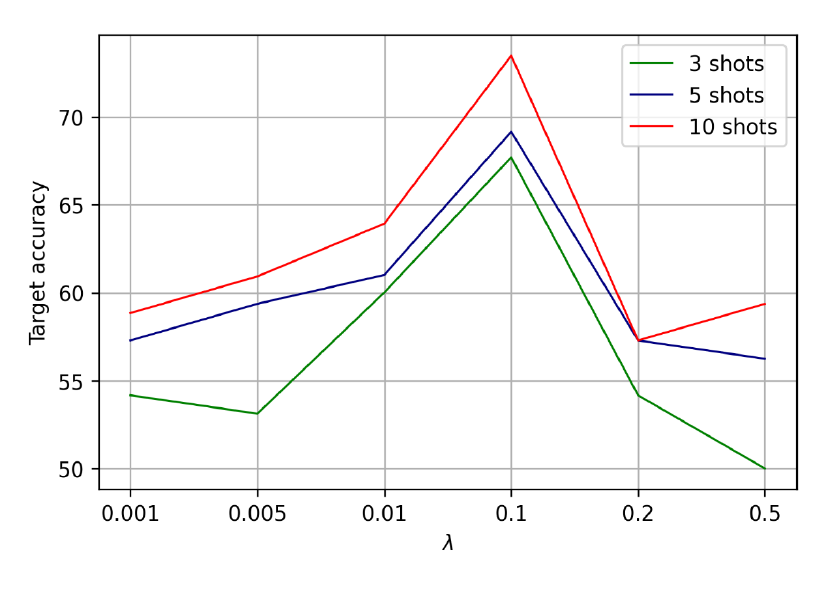} \\
\includegraphics[width=10cm]{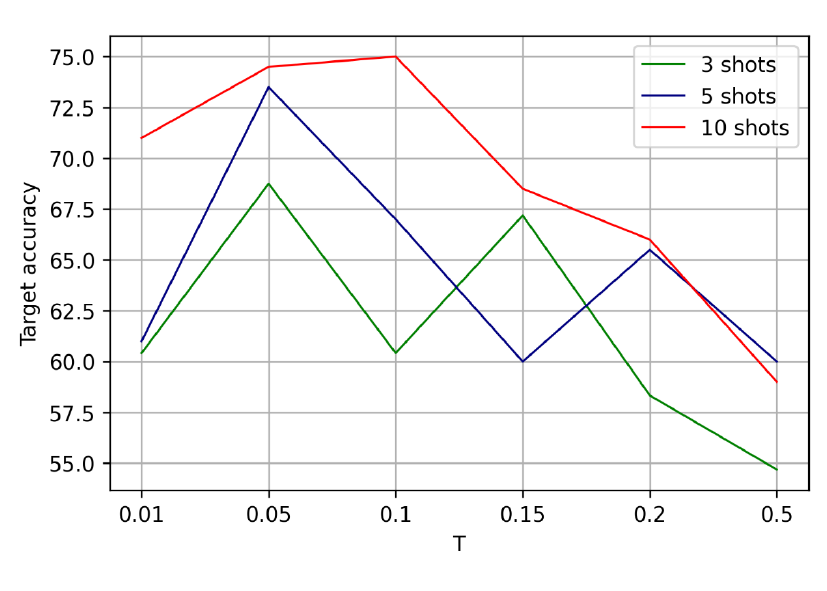} 
\end{center}
\caption{Ablation study. (a) Effect of parameter $\lambda  $  on target accuracy. (b) Effect of parameter $T$ on target accuracy.}       
\label{fig12}       
\end{figure}

Currently, $\lambda$ and $T$ are fixed to their optimal values and the model is trained using the four DA scenarios with different values of $K=3$, $5$, and $10$. Thus, a total of $12$ experiments are performed in order to plot the average loss and accuracy curves as given Figure~\ref{fig11a} and Figure~\ref{fig11b}, respectively. 

\begin{figure}[!ht]
\begin{center}
\includegraphics[width=11cm]{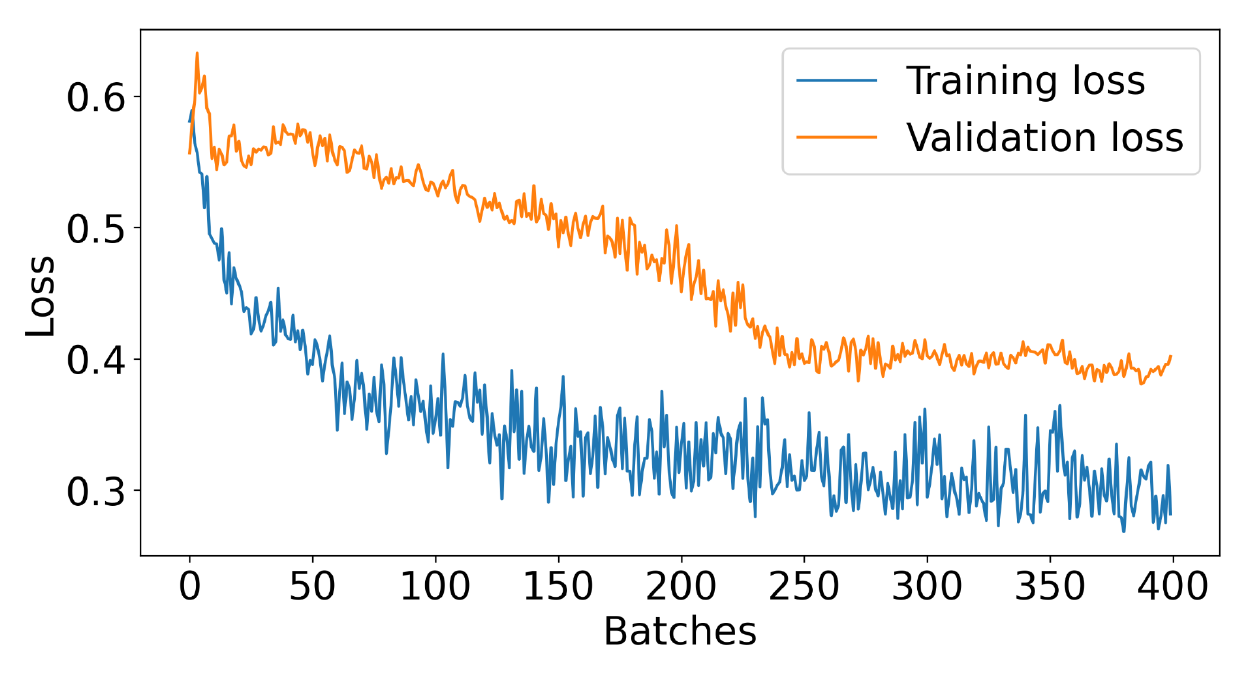} 
\end{center}
\caption{Average train and validation loss curves for source: SARS-CoV-2-CT and target: COVID19-CT-test. }
\label{fig11a}       
\end{figure}

\begin{figure}[!ht]
\begin{center}
\includegraphics[width=11cm]{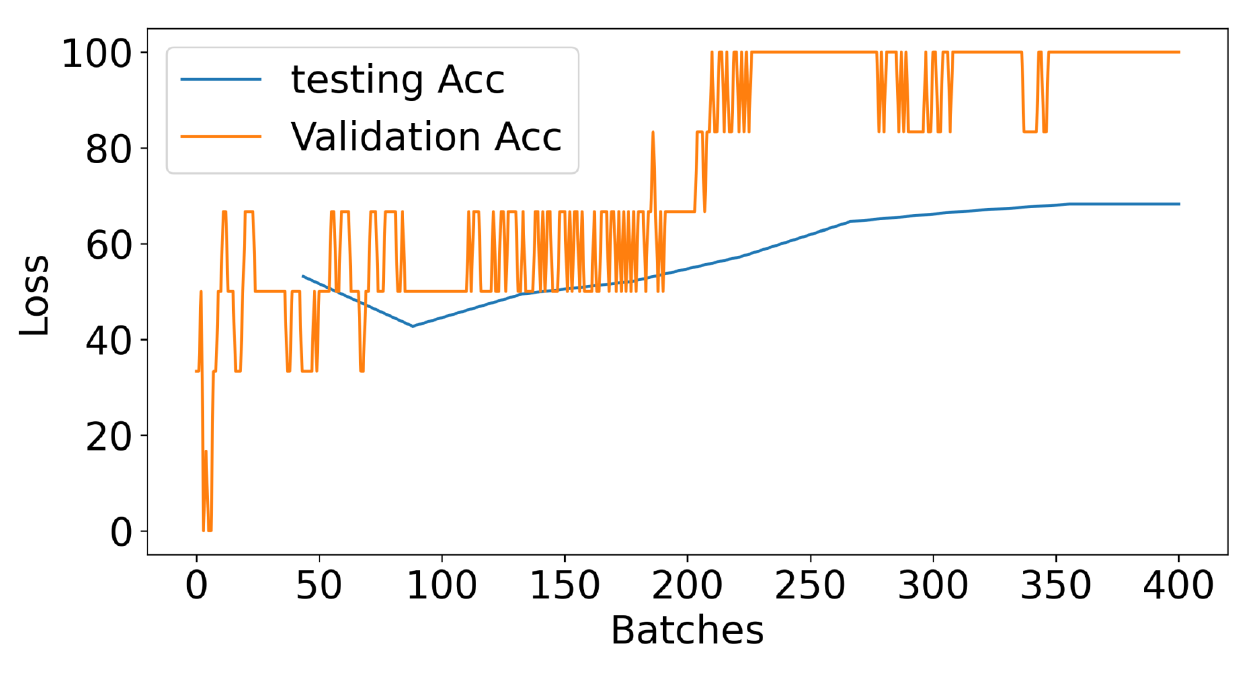} 
\end{center}
\caption{Average validation and test accuracy curves for source: SARS-CoV-2-CT and target: COVID19-CT-test. }
\label{fig11b}       
\end{figure}

As shown in Figure~\ref{fig11a}, the loss curve is not smooth. However, it has a general slow convergence towards zero. In addition, it is already converged to a stable value for 400 batches. The validation accuracy, shown in Figure~\ref{fig11b}, presents a high variability. However, it still can be used to get the best model (high validation accuracy) that is expected to correlate with the highest accuracy on the testing set.

Finally, a performance comparison is established between our proposed method and previous work done by Silva et al.~\cite{silva_covid-19_2020}. This work attempted cross-dataset classification in the field of COVID-19 detection based on CT scans (no other relevant research was found). In this work, they have simply trained a CNN model on the source dataset mixed with five labeled samples from the target dataset. Then, they tested the trained model on the target dataset. They have obtained the results shown in Table~\ref{tab10:final_results}. 

%
\begin{table}[!ht]
\caption{Comparison of COVID19-DANet to previous state-of-the-art. Samples per class refer to the number of labeled samples per class in the target dataset. The source dataset samples are all labeled.}
\label{tab10:final_results}       
%
%
\begin{center}
\begin{tabular} {p{2.5cm}p{2.5cm}cccc}    
    &           &  & \multicolumn{3}{c}{\textbf{COVID19-DANet {[}ours{]}}} \\ \cline{3-6} 
\textbf{}               & \multicolumn{1}{l}{\textbf{}} & \multicolumn{1}{l}{\textbf{}}             & \multicolumn{3}{c}{\textbf{samples per class}}                   \\ 

\textbf{Source dataset} & \textbf{Target dataset} &     \textbf{Silva et al.~\cite{silva_covid-19_2020} }                  & \textbf{3}                               & \textbf{5}                               & \textbf{10}                               \\ \hline
\textbf{SARS-CoV-2-}     & \textbf{COVID19-CT}                &                       &                                          &                                          &                                            \\ 
CT-scan                 & (Train)  & 59.12\%               & 62.22\%                                  & 63.52\%                                  & 66.11\%                                      \\  \hline

\textbf{SARS-CoV-2-}     & \textbf{COVID19-CT}   &                       &                                          &                                          &                                                   \\ 
CT-scan                 & (Test)            & 56.16\%               & 67.71\%                                  & 69.17\%                                  & 73.50\%                                   \\ \hline

\textbf{SARS-CoV-2-}     & \textbf{COVID19-CT}    &                       &                                          &                                          &                                                        \\ 
CT-scan                 & (Train + Test)     & 58.31\%               & 61.42\%                                  & 62.48\%                                  & 63.71\%                            \\ \hline
\textbf{COVID19-CT}     & \textbf{SARS-CoV-2-}      &                       &                                          &                                          &            \\  
(Train + Test)          & CT-scan         & 45.25\%               & 56.03\%                                  & 60.12\%                                  & 65.17\%                                          \\   \hline                            

\end{tabular}
\end{center}
\end{table}

As given in Table~\ref{tab10:final_results}, our proposed method  significantly outperforms the previous method. However, the results are still low, and there is room for improvement with other more suitable contributions. The method is able to remove a part (not all) of the data shift between source and target datasets. Especially, when $K=3$, the number of labeled target samples is just too small to provide a good representation. The same problem is encountered in the validation set, which is composed of $K$ labeled samples from each target class. Even though the validation accuracy reaches 100\%, the actual testing accuracy lags way behind. The introduced entropy loss function helped the model learn from the unlabelled samples. However, it has its limitations because minimizing the entropy on unlabelled data does not mean they are correctly classified. The model can still produce probability predictions for unlabelled target data that are very close to zero or one, even though the data is classified incorrectly.

\section{Conclusion}
\label{conclusion}

In this chapter, we proposed a method for the diagnosis of COVID-19 infection through the classification of CT images of the lung. Due to the limited data on COVID-19, we present a solution to tackle the classification problem of the cross-dataset through a DA technique. Our proposed method borrows ideas from few-shot learning by adding a prototypical layer on top of the feature extraction backbone, which is the pre-trained Efficientnet-B3 model without the top layer. We also proposed a combined loss function that is composed of the standard cross-entropy loss for class discrimination and another entropy loss computed over the unlabelled target set only. This so-called unlabelled target entropy loss is minimized and maximized in an alternative fashion, to reach the two objectives of class discrimination and domain invariance. 
The proposed solution has been tested with four different DA scenarios using the SARS-CoV-2-CT and COVID19-CT datasets. The achieved results have outperformed the state-of-the-art work in this field.

In our future work, we can improve performance by using other DA techniques using GAN-based approaches. In addition, we believe that techniques learning more from the unlabelled data are necessary to achieve better results. One interesting research direction consists of incorporating techniques of self-training and self-learning into the proposed DA method. These can help the model learn how to extract highly discriminative features from the unlabelled data directly.

\hspace{0.3cm}

\textbf{Acknowledgement}
This research project was supported by a grant from the Research Center of the College of Computer and Information Sciences, Deanship of Scientific Research, King Saud University.

\end{sloppy}

\bibliographystyle{spmpsci}
\bibliography{references}

\end{document}